%% file: ms.tex
\documentclass[letterpaper,12pt,notitlepage]{article}
\pdfoutput=1
\input{./TeX_files/preamble.arxiv}
\input{./TeX_files/preamble}
\newcommand{\nowordcount}[1]{#1}
\newcommand{\wordcount}[1]{}

\DeclareBibliographyCategory{cited}
\AtEveryCitekey{\addtocategory{cited}{\thefield{entrykey}}}
\bibliography{./ms}

\begin{document}
\pgfplotsset{
    every axis legend/.append style={
        at={(0.5,1.1)},
        anchor=south,
        line width=2pt
    }
}


\include{./data/wages_data}
\input{./shortened/title}

\nowordcount{\maketitle}
\input{./shortened/abstract}
\input{./shortened/introduction}
\input{./shortened/labor-equilibrium}
\input{./shortened/discussion}
\input{./shortened/policy}
\input{./shortened/conclusion}

\clearpage
\printbibliography
\end{document}

%% file: TeX_files/preamble.arxiv.tex
\usepackage[style=apa]{biblatex}
\usepackage[margin=1in]{geometry}

%% file: TeX_files/preamble.tex
\usepackage[utf8]{inputenc}
\usepackage{hyperref}
\usepackage{prettyref,varioref}
\usepackage{units}
\usepackage{amsmath,amssymb}

\usepackage{tikz}
\usepackage{pgfplots}
\usepackage{pgfplotstable}
\usepackage{pgfmath}
\usetikzlibrary{arrows,shapes.geometric,shapes.gates.logic.US,automata,calc,positioning}

\newrefformat{fig}{Figure \ref{#1} \vpageref{#1}}
\newcommand{\percapita}[1]{\nicefrac{#1}{C}}

%% file: data/wages_data.tex


\pgfplotstableread{./data/wages.dat}{\wagesdata}

\pgfplotstablegetelem{0}{cpiu}\of{\wagesdata}
\pgfmathsetmacro\basecpi{\pgfplotsretval}
\pgfmathsetmacro\basecpiurs{376.5}

\pgfplotstablecreatecol[expr={\thisrow{minw} / \thisrow{gdp_c}}]{Wmin}{\wagesdata}
\pgfplotstablecreatecol[expr={\thisrow{meanw} / \thisrow{gdp_c}}]{Wmean}{\wagesdata}
\pgfplotstablecreatecol[expr={\thisrow{medw} / \thisrow{gdp_c}}]{Wmed}{\wagesdata}
\pgfplotstablecreatecol[expr={\thisrow{minw} / \thisrow{meanw}}]{min_mean}{\wagesdata}
\pgfplotstablecreatecol[expr={\thisrow{minw} / \thisrow{medw}}]{min_med}{\wagesdata}
\pgfplotstablecreatecol[expr={\thisrow{medw} / \thisrow{meanw}}]{med_mean}{\wagesdata}
\pgfplotstablecreatecol[expr={\thisrow{min_med} / \thisrow{min_mean}}]{mmed_mmean}{\wagesdata}
\pgfplotstablecreatecol[expr={\thisrow{Wmin} / \thisrow{min_med}}]{min_med_gap}{\wagesdata}
\pgfplotstablecreatecol[expr={\thisrow{minw} / \thisrow{cpiurs}) * \basecpiurs}]{realminw}{\wagesdata}
\pgfplotstablecreatecol[expr=\thisrow{meanw} / \thisrow{cpiurs}) * \basecpiurs]{realmeanw}{\wagesdata}
\pgfplotstablecreatecol[expr=2080*\thisrow{avgnonsupwg}]{nonsupw}{\wagesdata}
\pgfplotstablecreatecol[expr=\thisrow{minw} / \thisrow{nonsupw}]{min_nonsup}{\wagesdata}

\pgfplotstablecreatecol[expr=52*\thisrow{medw_bls} / \thisrow{gdp_c}]{medw_bls_a}{\wagesdata}

\pgfplotstablecreatecol[expr={(\thisrow{medw_epi} * \thisrow{cpiurs} / \basecpiurs) * 2080 / \thisrow{gdp_c}}]{Wmed_epi}{\wagesdata}
\pgfplotstablecreatecol[expr={(\thisrow{meanw_epi} * \thisrow{cpiurs} / \basecpiurs) * 2080 / \thisrow{gdp_c}}]{Wmean_epi}{\wagesdata}
\pgfplotstablecreatecol[expr={\thisrow{Wmin} / \thisrow{Wmean_epi}}]{min_mean_epi}{\wagesdata}
\pgfplotstablecreatecol[expr={\thisrow{Wmin} / \thisrow{Wmed_epi}}]{min_med_epi}{\wagesdata}
\pgfplotstablecreatecol[expr={\thisrow{medw_epi} / \thisrow{meanw_epi}}]{med_mean_epi}{\wagesdata}

\pgfplotstablecreatecol[expr=\thisrow{min_med_epi}]{mmde}{\wagesdata}
\pgfplotstablecreatecol[expr=\thisrow{min_mean_epi}]{mmne}{\wagesdata}
\pgfplotstablecreatecol[expr=\thisrow{med_mean_epi}]{mdmne}{\wagesdata}
\pgfplotstablecreatecol[expr=2080 * \thisrow{medw_epi}]{realmedw}{\wagesdata}

\pgfplotstablecreatecol[expr={\thisrow{medw_epi} * 2080}]{real_medw_epi}{\wagesdata}

\pgfplotstablecreatecol[expr={\thisrow{cpiurs} / \prevrow{cpiurs} - 1}]{dcpiurs}{\wagesdata}

\pgfplotstablecreatecol[expr={\thisrow{gini} > 0 ? 1 - \thisrow{gini} : \thisrow{gini}}]{ngini}{\wagesdata}


\pgfplotstableread{./data/wages-gdp.dat}{\gdpdata}

\pgfplotstablecreatecol[expr={\thisrow{minw} / \thisrow{gdp_c}}]{Wmin}{\gdpdata}




\pgfplotstableread[col sep=tab,trim cells]{./data/wages-country-2018.dat}{\datanationE}

\pgfplotstablecreatecol[expr={\thisrow{national_minw_hr} * \thisrow{work_wk_hr} * 52 / \thisrow{national_gdp_c}}]{national_Wmin}{\datanationE}

\pgfplotstablecreatecol[expr={\thisrow{median_hh_income} / \thisrow{national_gdp_c}}]{med_hhi_gdpc}{\datanationE}

\pgfplotstableread{./data/gini-coefficient.dat}{\gini}

\pgfplotstablecreatecol[expr={1 - \thisrow{gini}}]{gini_inv}{\gini}

%% file: shortened/title.tex
\title{Minimum Wage, Labor Equilibrium, and the Productivity Horizon: A Visual Examination}
\author{John R. Moser\thanks{Moser: john.moser@ubalt.edu}}
\date{\today}

%% file: shortened/abstract.tex
\begin{abstract}
    In this paper, I present a visual representation of the relationship between mean hourly total compensation divided by per-capita GDP, hours worked per capita, and the labor share, and show the represented labor equilibrium equation is the definition of the labor share. I also present visual examination of the productivity horizon and wage compression, and use these to show the relationship between productivity, available employment per capita, and minimum wage.  From this I argue that wages are measured in relation to per-capita GDP, and that minimum wage controls income inequality and productivity growth.
\end{abstract}

%% file: shortened/introduction.tex
Literature and political dialogue on income inequality often focuses on real wages and inflation, with some recent shifting of focus from an inflation index to the Kaitz index for analyzing minimum wage.  Measures of income and earnings inequality include the Gini coefficient, labor share, and the 90/50 or 50/10 comparisons, all of which factor out the GDP deflator.  Little exploration has been done into wages as a nominal scalar with relationship to the per-capita GDP (\percapita{GDP}).

In this paper, I present and visualize labor share in terms of mean hourly total compensation, \percapita{GDP}, and per-capita hours worked, and show that minimum wage policy indirectly and strongly affects the mean hourly total compensation as a portion of \percapita{GDP}.  I also present visually the conditions under which labor substitution gains positive utility by technical progress and increases in total hourly compensation for low-productivity workers, which I term the ``productivity horizon''; and from this derive, in visual presentation, wage compression as the same function.

I argue from this that in an economy of rational actors, wage compression implies the mean wage is a function of the minimum wage; and that as total compensation is largely wage and much of the non-wage component is either not legally mutable or politically difficult to reduce, the minimum wage measured as a portion of \percapita{GDP} must control productivity.  I also argue that increases in minimum wage are technically indistinct from improvements in labor-saving technology, implying minimum wage has no long-run impact on per-capita labor hours and thus employment rates.

The first section presents the proposed labor equilibrium equation, equivalent to the labor share definition, and diagrams showing the described labor share event horizon and wage compression.  The second discusses the literature and empirical evidence supporting the implications for earnings inequality, employment, and productivity.  The third discusses the implications for public policy.  The final section concludes.

%% file: shortened/labor-equilibrium.tex
\section{Compensation as a Portion of Per-Capita GDP}

Wages are often considered in the context of real wage or, more recently, the relationship to median wage.  The body of literature frequently analyzes wages in the context of real wage; while this measures purchasing power, it doesn't measure economic context.  The visible fixation on real wages as the primary indicator and on externalities affecting real wages is characteristic of much economic research.

Nominal-to-nominal measures, such as nominal wage divided by per-capita GDP, are equivalent to real-to-real measures:  real wage divided by per-capita real GDP would divide out the GDP deflator.  Measurement of wage in relation to \percapita{GDP} changes the context from wage as an absolute buying power to wage as a portion of per-person output, and analogously changes the policy context of minimum wage from a "living wage" to a per-person standard-of-living.

\subsection{A Labor Equilibrium Equation}
\prettyref{fig:labor-equilibrium-equation} presents a form of the labor share definition as a labor equilibrium equation.  The term $T_{mean}$ is the mean hourly total compensation divided by \percapita{GDP}.  The equation itself is a refined form of an attempt to support the hypothesis that the ratio of minimum wage to \percapita{GDP} is the dominating factor in earnings inequality by its effect on the mean wage, the mathematical restrictions implied by the nature of GDP, and the phenomena of wage compression.

\nowordcount{
    \input{./shortened/figures/labor-share-gdpc-proof}
}

For the mean average total compensation for one labor-hour to remain a specific portion of \percapita{GDP}, a constant number of labor-hours worked per-capita requires the labor share to remain constant.  The diagram is merely a visualization of a division computation, particularly the equation shown.  This equation simplifies to the definition of labor share.

\begin{equation*}
    \begin{aligned}
        Labor\ Share & = \frac{T_{mean} \times Hours\ Worked}{C} \\
        Labor\ Share & = \frac{Total\ Hourly\ Compensation}{\percapita{GDP}}\times\frac{Hours\ Worked}{C} \\
        Labor\ Share & = \frac{Total\ Compensation\times C}{GDP \times Hours\ Worked}\times\frac{Hours\ Worked}{C} \\
        Labor\ Share & = \frac{Total\ Compensation}{GDP}
    \end{aligned}
\end{equation*}

According to the Bureau of Labor Statistics National Compensation Survey, wages and salaries since 1986 make up 70 to 73 percent of private industry total compensation; health insurance benefits—in many cases mandatory under the ACA—make up as low as 5.4 percent, ranging 7.0 to 7.8 percent since 2006; and legally-required social insurance payments make up 8 to 9 percent, totaling around 88 to 90 percent of total compensation.  This limits the amount of a wage increase employers can transfer to reduced benefits.

The bulk of what remains is paid leave, overtime and bonuses, and retirement benefits.  It seems unlikely employers will have the market power to significantly reduce paid leave or retirement benefits.  Further, low-wage workers are less-likely to receive defined benefits pensions, defer money to matched contributory benefits pensions, receive bonuses, and in some cases receive paid leave; even small minimum wage increases exceed what employers can offset by reducing other compensation.

This makes the mean wage as a portion of \percapita{GDP}, $W_{mean}=\frac{Mean\ Wage}{\percapita{GDP}}$, a viable proxy for $T_{mean}$.

\subsection{The Productivity Equilibrium Event Horizon}

A firm as a rational economic actor will use whatever method of production carries the least cost per good produced.  From the perspective that employers purchase, transform, and then sell inputs as goods, the two separate processes represent a ``transform unit''—interchangeable black boxes each accepting one unit of the exact same inputs and producing one unit of identical outputs—at different prices.  The price of these transform units represents the total cost of using high- versus low-productivity labor to produce a unit of output.

\prettyref{fig:productivity-equilibrium} shows two representations of this situation, assuming wage is the only difference in cost.  In practice, the whole cost difference to an individual firm includes rents and profits taken in the supply chain as well as wage—accounting for capital as a product of labor, such that firms involved in producing, shipping, and maintaining capital purchased and used by other firms are using labor to transform inputs to outputs, taking profits, and paying any rents on land or otherwise.

\nowordcount{
    \input{./shortened/figures/productivity-equilibrium-equation}
}

Both transform units have the same utility.  As long as the inequality remains the same, demand for high-productivity labor remains the same due to the availability of a lower-priced substitute.  When the relationship changes, technical progress crosses the productivity equilibrium horizon, on one side of which is the span of price relationships whereby it is more cost-effective to use low-productivity labor, and on the opposite more cost-effective to use high-productivity labor.

Improvements in technology reduce the cost of the high-productivity transform unit by reducing the labor-hours invested, which may reduce direct labor costs or increase the supply of capital at a lower price.  Increases in lower-productivity wages causes the same effect by increasing the price of the low-productivity transform unit, increasing the pace at which technical progress crosses the productivity equilibrium horizon and demand moves to high-productivity labor.

\subsection{Wage Compression}

\prettyref{fig:productivity-equilibrium-wage-compression} demonstrates increasing low-productivity wage and the corresponding increase in the cost of producing a good using low-wage labor.  This effect is called wage compression; I term negative wage compression ``sag'' as a descriptive shorthand.  The relationship of low-productivity labor costs versus substitution with high-productivity labor is unaffected by the method by which firms pay for labor costs.  Firms may transfer the cost of increased wages to consumers through price increases (a) or reallocate from profits to the labor share (c).

\nowordcount{
    \input{./shortened/figures/productivity-equilibrium-wage-compression}
}
We can measure the mean wage $W_{mean}$ in the same way as $T_{mean}$, being the mean hourly wage divided by the \percapita{GDP}.  Wage compression suggests we can describe $W_{mean}$ as a function of $W_{min}$, which I will call the ``Demand Function'' $\mathcal{D}$.  This suggest $W_{min}$ is a viable proxy to control $T_{mean}$.

Wage compression is well-explored in the current body of literature regarding minimum wage. Cunningham describes plainly, ``The lowest-paid workers have the greatest increases in wages when the minimum changes. Low-paid but above-the-minimum workers see their wages increase by a lesser amount, and workers at the top of the distribution do not see any changes in their wages. This results in compression of the wage distribution'' \autocite[41-42]{Cunningham2007}.  Engborn and Moser find evidence of wage compression in Brazil \autocite{Engborn2018}.  In Germany, introduction of a minimum wage significantly increased wages of low-wage workers relative to those of high-wage workers \autocite{Dustmann2020}.  Dube finds increases in family incomes caused by minimum wage increase diminish as income increases \autocite{Dube2019}.

%% file: shortened/figures/labor-share-gdpc-proof.tex
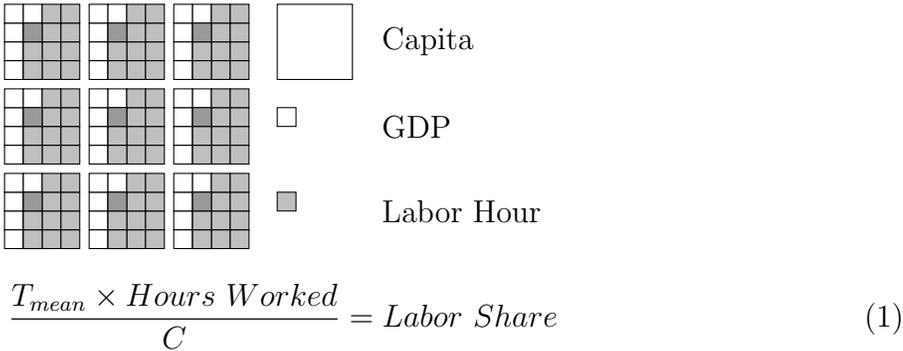
\begin{figure}[htb]
    \centering
    \begin{tikzpicture}[scale=0.25]
        \foreach \x in {0,...,2}
        \foreach \y in {0,...,2}
        {
            \foreach \xa in {0,...,1}
            \foreach \ya in {0,...,3}
            {
            \path [black, draw, fill=white]
                ({4.5 * \x + \xa},{4.5 * \y + \ya}) -- ++(0,1) -- ++(1,0) -- ++(0,-1) -- cycle;
            };
            \foreach \xa in {2,...,3}
            \foreach \ya in {0,...,3}
            {
                \path [black, draw, fill=gray!50]
                  ({4.5 * \x + \xa},{4.5 * \y + \ya}) --  ++(0,1) -- ++(1,0) -- ++(0,-1) -- cycle;
            };

            \foreach \xa in {1}
            \foreach \ya in {0,...,1}
            {
                \path [black, draw, fill=gray!50]
                ({4.5 * \x + \xa},{4.5 * \y + \ya}) --  ++(0,1) -- ++(1,0) -- ++(0,-1) -- cycle;
            };

            \foreach \xa in {1}
            \foreach \ya in {2}
            {
                \path [black, draw, fill=gray!80]
                ({4.5 * \x + \xa},{4.5 * \y + \ya}) --  ++(0,1) -- ++(1,0) -- ++(0,-1) -- cycle;
            };
        }

        \node (Legend) at ({4.5 * 3 + 1},{4.5 * 2}) {};
        \path [black, draw, fill=white]
         (Legend) -- ++(0,4) -- ++(4,0) -- ++(0,-4) -- ++(-4,0) -- cycle;
        \draw (Legend) ++(5,2) node[anchor=west] {Capita};

        \path [black, draw, fill=white]
        (Legend) ++(0,-2.5) -- ++(0,1) -- ++(1,0) -- ++(0,-1) -- cycle;
        \draw (Legend) ++(5,-2.5) node[anchor=west] {GDP};

        \path [black, draw, fill=gray!50]
        (Legend) ++(0,-7) -- ++(0,1) -- ++(1,0) -- ++(0,-1) -- cycle;
        \draw (Legend) ++(5,-7) node[anchor=west] {Labor Hour};
    \end{tikzpicture}
    \begin{equation} \label{eq:wmean-labor-share}
        \frac{T_{mean}\times Hours\ Worked}{C}={Labor\ Share}
    \end{equation}
    \caption{\label{fig:labor-equilibrium-equation}Labor equilibrium equation.}
\end{figure}

%% file: shortened/figures/productivity-equilibrium-equation.tex
\begin{figure}[htb]
    \centering
    \begin{tikzpicture}[scale=0.5]


        \path [draw=none]
        (-6,0) -- (6,0);

        \node at (0,1.5) {$\geq$};

        \foreach \x in {-4,...,-2,1,2,3}
        \foreach \y in {0,...,2}
        {
            \path [black, draw, fill=white]
            (\x,\y) -- ++(0,1) -- ++(1,0) -- ++(0,-1) -- cycle;
        };
        \foreach \x in {-3,...,-2}
        \foreach \y in {0,...,1}
        {
            \path [black, draw, fill=gray!50]
            (\x,\y) -- ++(0,1) -- ++(1,0) -- ++(0,-1) -- cycle;
        };

        \path [black, draw, fill=gray!75]
        (-2,2) -- ++(0,1) -- ++(1,0) -- ++(0,-1) -- cycle;

        \path [draw=none, fill=black]
        (1,0) ++(-0.25,0) ++(1,0) -- ++(0,2) -- ++(0.25,0) -- ++(0,0.25) -- 	++(2,0) -- ++(0,-2.25) -- cycle;
        \path [black, draw, fill=gray!50]
        (1,0) ++(3,0) -- ++(0,2) -- ++(-2,0) -- ++(0,-2) -- cycle;

        \node (Legend) at (-5,-2) {};

        \path [black, draw, fill=white]
        (Legend) ++(0,0) -- ++(0,1) -- ++(1,0) -- ++(0,-1) -- cycle;
        \draw (Legend) ++(1,0.5) node[anchor=west] {GDP};

        \path [black, draw, fill=gray!50]
        (Legend) ++(5,0) -- ++(0,1) -- ++(1,0) -- ++(0,-1) -- cycle;
        \draw (Legend) ++(6.5,0.5) node[anchor=west] {Labor Hour};

    \end{tikzpicture}
    \begin{equation} \label{eq:productivity-equilibrium}
        Wage_{0}\times Labor_{0}\geq Wage_1\times Labor_1
    \end{equation}
    \begin{equation} \label{eq:productivity-equilibrium-2}
        \frac{\text{Wage}_0}{\text{Wage}_1}\geq\frac{\text{Labor}_1}{\text{Labor}_0}
    \end{equation}
    \caption{\label{fig:productivity-equilibrium}Labor substitution to reach productivity equilibrium.  Each large block represents the GDP of an identical output good; shaded areas represent two equivalent ``transform units'' using different amounts of labor at different per-hour labor cost.  Black area in right diagram shows possible expansion of wages paid per labor-hour for high-productivity labor.}
\end{figure}

%% file: shortened/figures/productivity-equilibrium-wage-compression.tex
\begin{figure}[htb]
    \centering
    \begin{tikzpicture}[scale=0.35]


        \pgfmathsetmacro{\spacing}{12};

        \foreach \x in {-1,1}
        {
            \draw [black]
            (\spacing * \x / 2, 1.5) -- ++(0,-8);
        };

        \node at (-\spacing,-10) {(a)};
        \node at (0,-10) {(b)};
        \node at (\spacing,-10) {(c)};

        \foreach \y in {0,...,2}
        \foreach \x in {-3,...,-1}
        {
            \foreach \zz in {0,...,1}
            {
                \foreach \c in {0,1}
                {
                    \pgfmathsetmacro{\z}{(8/9) * (\zz + 2)/3)};
                    \path [black, draw, fill=white]
                    ({-\spacing + \x * \z}, {\z * \y - 4*\zz}) ++(\z - 1,0) -- ++(0,\z) -- ++(-\z,0) -- ++(0,-\z) -- cycle;

                    \pgfmathsetmacro{\z}{(8/9) * (\zz + 2)/3) + (1-0.8*\zz)/6};
                    \path [black, draw, fill=white]
                    ({\x * \z}, {\z * \y - 4*\zz}) ++(\z - 1,0) -- ++(0,\z) -- ++(-\z,0) -- ++(0,-\z) -- cycle;
                };

                \path [black, draw, fill=white]
                ({\x + \spacing}, {\y - 4*\zz}) ++(0,0) -- ++(0,1) -- ++(-1,0) -- ++(0,-1) -- cycle;
            };

            \foreach \z in {-1,...,1}
            {
                \path [black, draw, fill=white]
                ({\x + \spacing*\z}, {\y - 8}) ++(0,0) -- ++(0,1) -- ++(-1,0) -- ++(0,-1) -- cycle;
            };
        };

        \foreach \c in {-1,0,1}
        {
            \foreach \y in {0,...,2}
            \foreach \z in {0,...,2}
            {
                \foreach \x in {1,...,3}
                {
                    \path [black, draw, fill=white]
                    ({\x + \spacing * \c},\y) ++(0,{-4 * \z})-- ++(0,1) -- ++(1,0) -- ++(0,-1) -- cycle;
                };

            };
            \foreach \x in {-3,...,-2}
            \foreach \y in {0,...,1}
            {
                \path [black, draw, fill=gray!50]
                ({\spacing * \c + 2 * \x / 3 - 1/3},{2 * \y / 3}) -- ++(0,{2/3}) -- ++({2/3},0) -- ++(0,{-2/3}) -- cycle;

                \path [black, draw, fill=gray!50]
                (\spacing * \c + \x,\y) ++(0,-4) -- ++(0,1) -- ++(1,0) -- ++(0,-1) -- cycle;

                \path [black, draw, fill=gray!50]
                ({\spacing * \c + 9 * \x / 8 - 1/8},{9 * \y / 8}) ++(1/4,-8) -- ++(0,{9/8}) -- ++({9/8},0) -- ++(0,{-9/8}) -- cycle;
            };

            \node at ({\spacing * \c},1.5) {$\ngeq$};
            \node at ({\spacing * \c},{1.5-4}) {$=$};
            \node at ({\spacing * \c},{1.5-8}) {$>$};
            \path [black, draw, fill=gray!75]
            ({\spacing * \c + 1},-8) ++(-0.25,0) ++(1,0) -- ++(0,2.25) -- ++(2.25,0) -- ++(0,-2.25) -- cycle;
            \foreach \y in {0,...,1}
            {
                \path [black, draw, fill=gray!50]
                ({\spacing * \c + 1},{-4 * \y}) ++(3,0) -- ++(0,2) -- ++(-2,0) -- ++(0,-2) -- cycle;
            };
            \path [draw=none, fill=gray!50]
            ({\spacing * \c + 1},-8) ++(3,0) -- ++(0,2) -- ++(-2,0) -- ++(0,-2) -- cycle;
            \path [black, draw]
            ({\spacing * \c + 1},-8) -- ++(3,0) -- ++(0,3);
        };

    \end{tikzpicture}
    \begin{equation} \label{eq:demand-function}
        W_{mean}=\mathcal{D}\left(W_{min}\right)
    \end{equation}
    \caption{\label{fig:productivity-equilibrium-wage-compression}Wage compression illustrated.  Minimum wage increases by 68.75 percent; the cost of using low-productivity labor increases to 25 percent more than using high-productivity labor.  Wage increases absorbed by price increases (a), reduced profits (c), or a mix (b).}
\end{figure}

%% file: shortened/discussion.tex
\section{Discussion}

Unlike measurement by mean wage, measurement in terms of share of \percapita{GDP} accounts for changes in the economic context.  Literature has used analysis of wage bins to measure lost and excess jobs after an increase in real minimum wage increase and determine the impact on overall employment \autocite{Cengiz2019}.  Would that productivity increased more than real minimum wage in the period analyzed, there would be a net reduction in wages as a portion of \percapita{GDP}.  Such a situation is unlikely, e.g. in Hungary the doubling of nominal minimum wage between 2000 and 2002 raised the annual minimum wage for a 2,080-hour work year from .406 to .606 times \percapita{GDP} concurrent with large increases in capital stock \autocite{Harasztosi2019}.

The relationships suggested here lend support to the argument that minimum wage—specifically, the relationship of minimum wage to per-capita GDP—is the dominating factor, and that low $W_{min}$ leads to sag in the wage structure, causing lower labor share, reduced productivity growth rates, and increased income inequality.

\subsection{Evidence for Wage and Earnings Inequality}

\prettyref{fig:minimum-gdp-union-pr} provides an empirical view of the United States, computing $W_{min}$ and $W_{mean}$ as annual approximations by multiplying the hourly wage by 2,080 hours. \prettyref{fig:Minimum-Mean-Wage-Scatter} further examines the relationship between $W_{min}$ and $W_{mean}$.  Minimum wage only fell below 45 percent after 1983, and afterwards values close to 31 percent repeatedly occur as early as 1988 and as late as 2010, while the span from 1987 to 1997 repeatedly crosses the range from 30 to 35 percent.

More directly, \prettyref{fig:minimum-gini} shows the Gini coefficient rapidly rising as minimum wage falls, and holding generally level as minimum wage remained level.  Gini coefficient data is not generally published annually, so is lower resolution.

While the relationship between minimum wage, mean wage, and \percapita{GDP} is clearly visible in Figure \ref{fig:minimum-gdp-union-pr}, union participation rate continues to fall even after $W_{min}$ and $W_{mean}$ change to a more shallow trend.  The minimum wage established in Germany, which at the time had never had a minimum wage policy in its history, was €8.50 hourly, and higher than many union-negotiated wages \autocite{Dustmann2020}, suggesting union wage-setting power is built on minimum wage.  While unions have an effect on wage negotiations, their ability to shift the demand curve is limited.

Although many employers indicate they would respond to minimum wage increases by slowing pay increases for higher-skilled employees \autocite{Schmitt2013}, the data show higher minimum wages lead to higher mean wages when measured in relation to \percapita{GDP}.  Likewise, such delays would accelerate movement across the productivity equilibrium horizon, increasing demand for higher-wage workers; and wage compression would only be avoided by accelerating pay increases for higher-skilled employees.

Wage compression is weaker when $W_{min}$ is higher:  as wages compress, changes of the same degree in $W_{min}$ represent larger portions of $W_{mean}$, increasing the degree to which $W_{mean}$ can increase without crossing back over the productivity equilibrium event horizon.  Marginal increases in $W_{min}$ generate higher marginal increases in $W_{mean}$ and, we would expect, smaller marginal increases in productivity as $W_{min}$ increases.

Price effects are also minimal, allowing minimum wage increases to effectively reduce poverty for low-wage workers.  In the case of Hungary, prices were 20 percent higher across those two years; ultimately, the real minimum wage increased by 65 percent, the ratio to median wage by 58 percent, and the ratio to \percapita{GDP} by 49 percent.  Studies in the United States find a 10 percent increase in minimum wage leads to roughly a half percent of increase in overall prices \autocite{Schmitt2013}.  Gramlich observes the 25 percent increase in 1974 ``raise the wage bill directly by only 0.4 percent and indirectly by another 0.4 percent'' \autocite[430]{Gramlich1976}.

\subsection{Evidence for Employment}

Were firms to distribute exactly 100 percent of the increased wage costs to price, the ratio of total compensation to \percapita{GDP} would increase:  the proportional increase in wage would be greater than the proportional increase in price.  This increase in incomes increases aggregate demand, which increases demand for labor.  Harasztosi and Lindner find the majority of cost increase from a near-doubling of minimum wage in Hungary shifted to prices, with only a 10.8 percent price increase in the long term despite only 25 percent of the wage costs shifting from firms to the labor share \autocite{Harasztosi2019}.

The strong correlation between $W_{min}$ and $W_{mean}$, and thus $T_{mean}$, seen in \prettyref{fig:Minimum-Mean-Wage-Scatter} also indicates increased labor demand:  a higher labor share with the same hours worked per capita only occurs by an increased $T_{mean}$, and so more labor must enter the market at wages below the expected $T_{mean}$ to restore equilibrium.  Although the demand function $\mathcal{D}$ is presumed from an observation and not a mathematical relationship, the evidence does suggest this equilibrium exists.


Sharpe, Arsenault, and Harrison construct a similar model showing the relationship between productivity, total compensation, and labor share.  They relate \percapita{GDP} to labor productivity as ``the product of labour productivity […] average hours worked per employed person, and the employment/population ratio'' \autocite[19]{Sharpe2008}, finding labor productivity to account for over 75 percent of \percapita{GDP} growth in Canada between 1961 and 2007.  Recent work also finds the amount of non-work done at work to be around 7 percent or 34 minutes, roughly half of which is spent eating \autocite{Burda2019}, suggesting labor productivity is almost entirely a matter of technological development.


As both labor productivity and minimum wage cause convergence on the productivity event horizon, we would expect a minimum wage to have long-term effects on productivity, but no long-term effect on employment.

This is consistent with a wide body of literature.  Cengiz, Dube, Lindner, and Zipper find little to no negative effects particularly on employment, notably in non-tradable sectors such as services, with the most notable impacts on tradable sectors \autocite{Dube2016,Cengiz2019}.  Dustmann, Lindner, et al find no reduction in employment prospects for low-wage workers nor increased likelihood of workers switching to new employers after the establishment of a minimum wage in Germany \autocite{Dustmann2020}.  Evidence from Hungary finds minimal disemployment effects despite the extreme rate of increase \autocite{Harasztosi2019}.  Engborn and Moser find only a modest affect on employment \autocite{Engborn2018}.

No reasonable purpose presents itself suggesting different market responses to crossing the productivity event horizon by improvements in labor-saving technology or changes in the relationship between wages.  If minimum wage increases caused a reduction in labor demand, it could only be explained by shifting the same GDP from low-productivity labor to less high-productivity labor.  This is exactly the situation in which technical progress occurs without rising wages, and so the long-term employment impacts must be the same.  As employment demand is stimulated by consumer spending, any increases in unemployment by such technical progress would appear to be a money shortage, and resolved by increasing the money supply.  Such measures are not generally concurrent policy when increasing minimum wage, and yet unemployment rates do not appear to respond to minimum wage increases.

\subsection{Evidence for Productivity}

Much literature supports the expectation of productivity growth from the theoretical basis given here.  Evidence in Brazil from a 119 percent real minimum wage increase from 1996 to 2012 finds a relocation of workers to more-productive businesses \autocite{Engborn2018}.  Workers in Germany found higher-paying, more-productive employment after the establishment of its first-ever minimum wage, and were unlikely to move from region to region to find new jobs \autocite{Dustmann2020}.  Rather, regions in which a higher portion of employment was in low-wage jobs experienced greater increases in productivity than those where wages were already generally high:  better-paying jobs came to the workers.

It seems intuitive that an extreme wage increase would more likely cause adverse effects than a slower increase.  Factories and machines take time to build, and an increase in demand for these leads to a shortage, forcing employers to shift the cost to increased prices.  Despite this, examination of a 57 percent minimum wage increase in Hungary from 2000 to 2001, and by a total of 96 percent from 2000 to 2002 \autocite{Harasztosi2019}, finds large increases in capital stock, indicating an increase in the use of labor-saving technology and thus productivity.  This increase amounts to a change across two years from .406 to .606 times \percapita{GDP} for a full-time work year; like the US, Hungary has a 40 hour work week.  Assuming an extreme rate of increase does cause greater negative effects, what counts as ``extreme'' is apparently quite higher than is intuitive, at least in Hungary at that time.

Dube suggests increased wages are borne by the middle and upper class through small price increases \autocite{Dube2019}.  A minimum wage increase in relation to \percapita{GDP} implies, due to the increase in higher wages in relation to \percapita{GDP}, that prices do not increase relative to wages and that middle- and upper-income earners enjoy an increase in purchasing power.  These two views do not necessarily conflict, as any productivity gains imply that prices increase more slowly than wages, which can ultimately result in adding more than is removed from an individual's purchasing power.

\nowordcount{
    \input{./TeX_files/figures/minimum_wage_gdpc_time_series}
    \input{./TeX_files/figures/minimum_wage_gdpc}
    \input{./TeX_files/figures/minimum_wage_gini_time_series}
}

%% file: TeX_files/figures/minimum_wage_gdpc_time_series.tex
\begin{figure}[pthb]
    \pgfplotsset{
        every axis legend/.append style={
            at={(1,1)},
            anchor=north east,
            line width=2pt
        }
    }

    \begin{tikzpicture}
        \begin{axis}
            [
                xlabel=Year,
                grid=major,
                legend entries=
                {
                    Minimum wage $\div$ $\percapita{GDP}$,
                    Mean wage $\div$ $\percapita{GDP}$,
                    Minimum wage $\div$ Mean wage,
                    {U3 Unemployment (right)},
                    {Union Participation (right)},
                    {Union Participation (CRS) (right)}
                },
                width=0.96*\textwidth,
                height = 0.80 * \textheight,
                x tick label style={
                    /pgf/number format/1000 sep={}
                },
                yticklabel={$\pgfmathparse{\tick*100}\pgfmathprintnumber\pgfmathresult\%$},
                xmin=1960,
                xmax=2018,
                ymin=.0,
                ymax=1.3
            ]

            \addplot
            [
                mark=none,
                x=Year,
                color=blue
            ]
            table
            [
                x=year,
                y=Wmin
            ] {\wagesdata};

            \addplot
            [
                mark=none,
                x=Year,
                color=olive
            ]
            table
            [
                x=year,
                y=Wmean
            ] {\wagesdata};


            \addplot
            [
                mark=none,
                x=Year,
                color=red
            ]
            table
            [
                x=year,
                y=min_mean
            ] {\wagesdata};

            \addlegendimage{mark=none, color=violet, dashed};
            \addlegendimage{mark=none, color=orange};
            \addlegendimage{mark=none, color=orange, dashed};

        \end{axis}
        \begin{axis}
        [
            grid=none,
            xmin=1960,
            xmax=2018,
            ymin=0.02,
            ymax=.28,
            hide x axis,
            axis y line*=right,
            width = 0.96*\textwidth,
            height = 0.8 * \textheight,
            yticklabel={$\pgfmathparse{\tick*100}\pgfmathprintnumber\pgfmathresult\%$},
            xtick pos=right,
            set layers=standard
        ]
            \addplot
            [
                mark=none,
                x=Year,
                color=orange
            ]
            table
            [
                x=year,
                y=upr_bls
            ] {\wagesdata};

            \addplot
            [
                mark=none,
                x=Year,
                color=orange,
                dashed
            ]
            table
            [
                x=year,
                y=upr_cuilr
            ] {\wagesdata};

            \addplot
            [
                dashed,
                mark=none,
                x=Year,
                color=violet
            ]
            table
            [
            x=year,
            y=u3unemp
            ] {\wagesdata};

            \legend{};
        \end{axis}

    \end{tikzpicture}

    \caption{\label{fig:minimum-gdp-union-pr}Minimum and mean wage as a percentage of per-capita GDP and each other.  Mean wage from the Social Security Administration \autocite{USSSACOLA2020}.  U3 unemployment and union participation on the right axis, from the Bureau of Labor Statistics and the Congressional Research Service \autocite{Mayer2004}. BLS Occupational Employment Survey data retrieved from the Economic Policy Institute \autocite{EPIData2019}.  GDP data from the World Bank \autocite{WorldBank}.}
\end{figure}

%% file: TeX_files/figures/minimum_wage_gdpc.tex
\begin{figure}[ptbh]
    \pgfplotsset{
        every axis legend/.append style={
            at={(0.5,1.1)},
            anchor=south,
            line width=2pt
        }
    }

    \begin{tikzpicture}
        \begin{axis}
            [
                xlabel=Minimum Wage $\div$ $\nicefrac{GDP}{C}$,
                grid=major,
                legend entries=
                {
                    Minimum wage $\div$ Mean wage,
                    Mean wage $\div$ $\nicefrac{GDP}{C}$,
                    Union Participation Rate (not to scale)
                },
                width=\textwidth,
                yticklabel={$\pgfmathparse{\tick*100}\pgfmathprintnumber\pgfmathresult\%$},
                xticklabel={$\pgfmathparse{\tick*100}\pgfmathprintnumber\pgfmathresult\%$},
                ymin=0,
                ymax=1.40
            ]

            \addplot
            [
                only marks,
                x=Year,
                color=red
            ]
            table
            [
                x=Wmin,
                y=min_mean
            ] {\wagesdata};

            \addplot
            [
                only marks,
                x=Year,
                color=olive
            ]
            table
            [
                x=Wmin,
                y=Wmean
            ] {\wagesdata};

            \addplot
            [
                only marks,
                mark=x,
                x=Year,
                color=blue
            ]
            table
            [
                x=Wmin,
                y expr={\thisrow{upr_cuilr}*2.75+.465}
            ] {\wagesdata};

            \addplot
            [
                mark=none,
                color=red!40
            ]
            table
            [
                x=Wmin,
                y expr=-0.791*\thisrow{Wmin}^2 + 1.35*\thisrow{Wmin} + 0.0273
            ] {\wagesdata};

            \addplot
            [
                mark=none,
                color=olive!40
            ]
            table
            [
                x=Wmin,
                y expr=0.537*\thisrow{Wmin}^2 + 0.378*\thisrow{Wmin} + 0.665
            ] {\wagesdata};
        \end{axis}
    \end{tikzpicture}

    \caption{\label{fig:Minimum-Mean-Wage-Scatter}$\frac{W_{min}}{W_{mean}}$ and $W_{mean}$ plotted against $W_{min}$ with quadratic regression.  Union participation rate is stretched onto a different Y axis (not shown) to illustrate impact on mean wage.}
\end{figure}
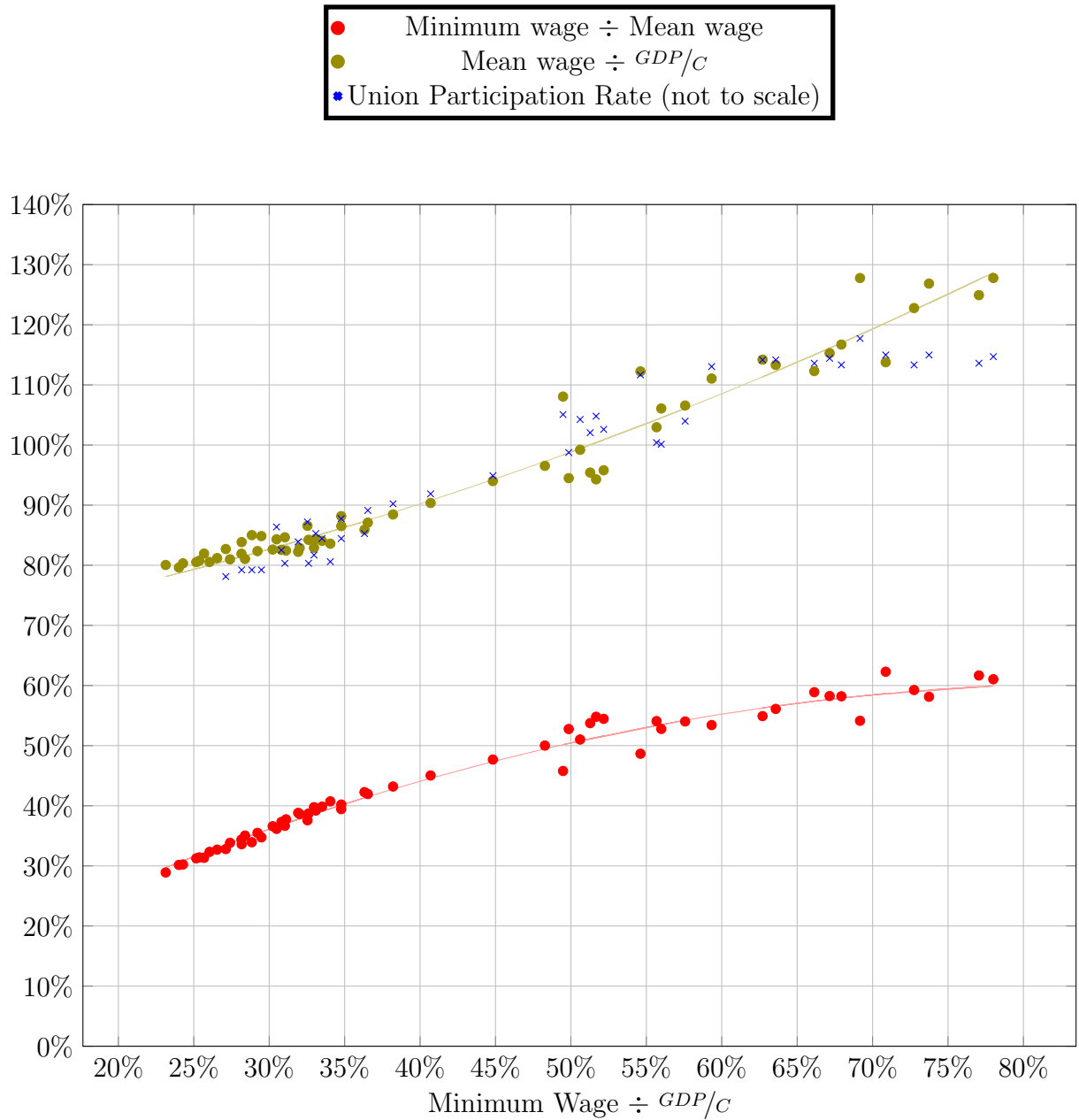

%% file: TeX_files/figures/minimum_wage_gini_time_series.tex
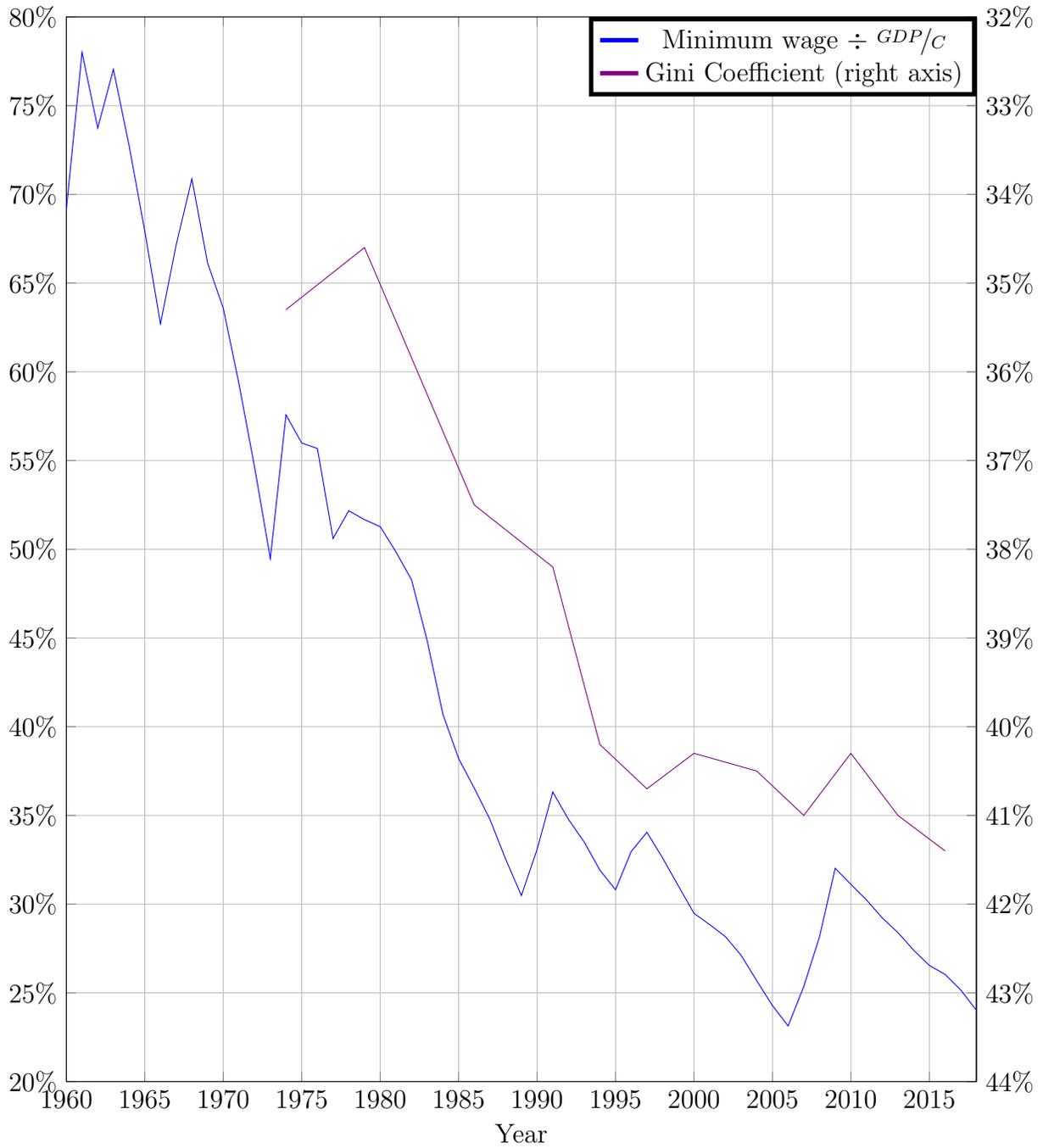
\begin{figure}[pthb]
    \pgfplotsset{
        every axis legend/.append style={
            at={(1,1)},
            anchor=north east,
            line width=2pt
        }
    }

    \begin{tikzpicture}
        \begin{axis}
            [
                xlabel=Year,
                grid=major,
                legend entries=
                {
                    Minimum wage $\div$ $\percapita{GDP}$,
                    {Gini Coefficient (right axis)}
                },
                width=0.96*\textwidth,
                height = 0.80 * \textheight,
                x tick label style={
                    /pgf/number format/1000 sep={}
                },
                yticklabel={$\pgfmathparse{\tick*100}\pgfmathprintnumber\pgfmathresult\%$},
                xmin=1960,
                xmax=2018,
                ymin=.20,
                ymax=.80
            ]

            \addplot
            [
                mark=none,
                x=Year,
                unbounded coords=jump,
                color=blue
            ]
            table
            [
                x=year,
                y=Wmin
            ] {\wagesdata};
            \addlegendimage{mark=none, color=violet};
        \end{axis}
        \begin{axis}
            [
                grid=none,
                xmin=1960,
                xmax=2018,
                ymin=0.32,
                ymax=0.44,
                y dir=reverse,
                hide x axis,
                axis y line*=right,
                width = 0.96*\textwidth,
                height = 0.8 * \textheight,
                    yticklabel={$\pgfmathparse{\tick*100}\pgfmathprintnumber\pgfmathresult\%$},
                xtick pos=right,
                set layers=standard
            ]
            \addplot
            [
                mark=none,
                x=Year,
                unbounded coords=jump,
                color=violet
            ]
            table
            [
                x=year,
                y=gini
            ] {\wagesdata};
        \end{axis}
    \end{tikzpicture}

    \caption{\label{fig:minimum-gini}Minimum wage as a percentage of per-capita GDP plotted against the Gini coefficient from the World Bank \autocite{WorldBank}.}
\end{figure}

%% file: shortened/policy.tex


\section{Policy Implications}

The analysis here indicates minimum wage is a critical and foundational economic policy required for the health of an economy.  without a minimum wage, wage compression operates in reverse—sag.  While aggregate supply and demand indicate a single price-quantity equilibrium, some suppliers are willing to supply below that equilibrium; this available lower-priced labor prohibits markets from raising wages.

Some workers are willing to accept a lower wage—for example if they need income for basic needs and have no other immediate options.  Firms hiring these workers at lower wages have lower, increasing supply.  Specifically, these firms are willing to supply more of a good than other firms at a lower price.  Consumers will purchase a good at the lowest price available to them, and the surplus reduces the equilibrium price.

This reduces demand for goods:  consumers shift their purchasing to firms employing lower-wage workers and supplying those goods at a lower price, and less quantity is purchased at the previous, higher prices.  Firms must charge prices sufficient to pay wages, so higher-wage firms are limited in how far they can reduce price.  The demand for labor falls, and a labor surplus occurs:  there are fewer job opportunities—less quantity of labor demanded—at the original, higher wage.

This indicates markets cannot permanently lift wages because price discovery will always find lower-wage labor so long as there are enough workers unable to find higher-paying jobs, and employers can better price-compete by paying lower wages.  The bottom wage must fall ever further behind productivity.  Minimum wage policy sets a price floor above the price where more than zero labor is supplied, preventing labor demand from falling.

Up to now, minimum wage policy proposals have focused on indexing to inflation  or the Kaitz index (median wage).  A 2016 report by the congressional Research Service examined indexes to various consumer price indices, personal consumption expenditure, employment cost index, average hourly earnings for manufacturing workers, and average hourly earnings for production and nonsupervisory workers \autocite{Brummund2016}.  These reflect various State policies or prior policy proposals.  Other recent proposals have focused on median wage or regional price parity \autocite{Dube2014,USCongress2019HR582}.

Regarding inflation indexing, using 2,080 hours of minimum wage and adjusting to CPI-U-RS, the real annual minimum wage in 1960 was \$2,080; in 1985 was \$2,090; in 2001 was \$2,047; and in 2017 was \$2,076.  For the same years, the real mean wage was \$3,842, \$4,838, \$6,032, and \$6,643; \prettyref{fig:minimum-gdp-union-pr} shows the relationship between mean and minimum wage over this whole span.  The difference between minimum and mean average wages increases in the long run for a level real minimum wage.

Pertaining to the Kaitz index, according to the Bureau of Labor Statistics OES reports, the ratio of minimum to median wage was 0.45 in 1985, 1992, 1999, 2011, and 2012.  During these years, the ratio of minimum to mean wage was 0.38, 0.37, 0.36, 0.35, and 0.34.  For 1989, 2002, and 2017, when the ratio of minimum to median was 0.40, the ratio of minimum to mean was 0.33, 0.31, and 0.30.  Although the decline is slower than with real wage, for the same range of minimum to median wage, the ratio of minimum to mean wage decreases in later years.

Pertaining to the ratio of minimum wage to average nonsupervisory wage, as proposed by the PHASE Act \autocite{USCongress2020HR2080}, the ratio was within 1.2 percentage points of 40 percent in 1972, 1983, 1984, 1991, 1992, 1996, 1997, 1998, and 2009.  Nonsupervisory wages were generally above 40 percent prior to 1983 and below in later years; as with general trends in other measures of wage, this is expected as an effect of wage compression.

The strong, direct, and reactive relationship between minimum wage and mean wage when measured as a portion of \percapita{GDP} suggests the relationship between wages and \percapita{GDP} is the measure of wage in its economic context.  Real wages measure a wage in the context of means and in comparison in that context to wages in other years, but not in the context of an economy.  Inflation is a valuable and important economic indicator; for example, real GDP is a consistent year to year measure of the productive output of an economy.  This consistency and appropriateness does not translate to the measurement of wages if we are to assume that a null change in minimum wage should result in no long-run wage compression, which seems a reasonable assumption if the definition of wage compression is that an increase in minimum wage causes wages to become closer together.

As such, if policy aims to control income inequality, and so must control earnings inequality, then it must index minimum wage to \percapita{GDP}.  An index target of $\nicefrac{2}{3}$ reflects some of the lower points between 1950 and 1970; the highest has been 109 percent of \percapita{GDP} in 1940.  The optimal value is outside the scope of this paper, and besides is best found by careful increases and monitoring of the economy; however, for consideration, per-capita measures use the whole population, which includes those not yet of age to join the workforce, those retired, and general non-participants, making the possible—but not necessarily efficient—targets quite high.

The evidence from Hungary provides interesting considerations as well.  One possible policy is to set the minimum wage to its current portion of \percapita{GDP} and increase that portion by 2.5 percent each year until reaching a target; yet Hungary achieved an increase of 20 percent of \percapita{GDP} in two years, or a 50 percent increase of the measure.  The evidence from Germany, regarding growth in regional productivity, suggests States implementing such policy will improve regional productivity relative to their neighbors, although this is best examined in research regarding GDP effects of minimum wage increases in U.S. States relative to states which didn't increase minimum wage in the same time frame.

Whether the United States would respond as Hungary to rapid increases in minimum wage is a question best answered by flexible policy allowing a regulatory board to raise minimum wage considerably rapidly, but also to quickly slow further increases if there are negative economic consequences.  The same is advisable for States regarding this and the observed effects of regional productivity growth, which may prove interesting to State policymakers seeking to increase their State's GDP without waiting for Federal action.

Finally, the productivity horizon effect indicates the long-standing puzzle of decreasing productivity growth rates is driven by a sagging wage structure due to a falling minimum wage.  The theory and evidence suggests indexing to \percapita{GDP} would stabilize this, while indexing to other measures would ensure continuing decline.


%% file: shortened/conclusion.tex
\section{Conclusion}

One conceptual explanation of a fair minimum wage is that there's a pie for each person, and the pies get bigger.  An inflation-indexed wage keeps the same size slice, but reduces the portion as the pie grows; while a \percapita{GDP} index represents keeping the same portion so the slice grows with the pie.  This metaphor lead to my original hypothesis that an index to \percapita{GDP} would control earnings inequality; and this paper carries that forward, notably in that the theoretical explanation of the mechanism behind wage compression didn't come to mind until I'd illustrated the process of crossing the productivity equilibrium horizon.  This method of searching for insights is  similar to economic thought experiments \autocite{Krugman1997} or mathematical visual proofs \autocite{RelafordDoyle2017}.

Here I have presented reasonable theoretical mechanisms for several findings in recent literature and long-term empirical observations.  Much of this is known, although the debate over whether minimum wage affects unemployment or productivity is largely answered at current by empirical studies and complex statistical analysis, and I found little explanation of a mechanism behind wage compression.  In providing a mechanism for minimum wages to increase productivity, I've observed that immediate minimum wage increases are functionally equivalent to advances in technology in terms of impacts on productivity and employment.

The implications are consistent with empirical literature demonstrating wage compression, productivity gains, and capital accumulation when minimum wages increase, as well as the evidence and literature regarding the impacts of minimum wage on employment.  The empirical evidence in the United States also shows minimum wage, when measured as a portion of per-capita GDP, has a dominating and consistent effect on the relationship of minimum wage to mean wage, suggesting minimum wage has a dominating effect on earnings inequality.

Finally, I have suggested policymakers seeking to prevent falling productivity gains and rising income and earnings inequality should set index minimum wage to per-capita GDP targets consistent with international and historic United States levels.  No other index appears to have this solidifying effect, and for good reason:  wages can't sag or compress when minimum wage is fixed to per-capita GDP, as that would require the per-capita GDP to become less or greater in relation to minimum wage, or for large parts of the wage bill to move in polarizing directions in a manner inconsistent with wage compression.


%% file: ms.bib
@InProceedings{RelafordDoyle2017,
  author    = {Josephine Relaford-Doyle and Rafael Nunez},
  booktitle = {CogSci 2017 Proceedings},
  title     = {When does a 'visual proof by induction' serve a proof-like function in mathematics?},
  eventdate = {2017-07-26},
  isbn      = {978-0-9911967-6-0},
  pages     = {1004--1009},
  url       = {https://cogsci.mindmodeling.org/2017/papers/0196/paper0196.pdf},
  year      = {2017},
}

@Report{Sharpe2008,
  author      = {Andrew Sharpe and Jean-François Arsenault and Peter Harrison},
  institution = {Center for the Study of Living Standards},
  title       = {The Relationship Between Labour Productivity and Real Wage Growth in Canada and OECD Countries},
  type        = {resreport},
  number      = {2008-8},
  url         = {http://www.csls.ca/reports/csls2008-8.pdf},
  month       = {12},
  year        = {2008},
}

@Book{Cunningham2007,
  author      = {Wendy Cunningham},
  date        = {2007-06-12},
  title       = {Minimum Wages and Social Policy: Lessons from Developing Countries},
  isbn        = {0-8213-7011-1},
  publisher   = {World Bank Publications},
  url         = {https://documents.worldbank.org/en/publication/documents-reports/documentdetail/826061468142780021/minimum-wages-and-social-policy-lessons-from-developing-countries},
  institution = {World Bank Group},
  year        = {2007},
}

@Report{Dustmann2020,
  author      = {Dustmann, Christian and Linder, Attila and Schönberg, Uta and Matthias, Umkehrer and vom Berge, Philipp},
  date        = {2020-02-20},
  institution = {Center for Research and Analysis of Migration (CReAM)},
  title       = {Reallocation Effects of the Minimum Wage},
  type        = {resreport},
  url         = {https://www.cream-migration.org/publ_uploads/CDP_07_20.pdf},
  urldate     = {2020-06-16},
  comment     = {Recommended by Prof. Notowidigdo in a response to a submission.  Purpose:  studies the potential productivity-enhancing effects of the minimum wage.

Page 3: "There is no indication that it lowered the employment prospects of low-wage workers"

Minimum wage didn't reduce employment in regions where more workers were affected (generally lower wage before minimum wage) "helped reducing wage inequality"

Low-wage workers at regional and individual level appear to reallocate to higher-wage jobs at higher-wage firms, while higher-wage workers are less likely to move.

Page 7: " within-plant productivity increases (Bossler, Gürtzgen, Lochner, Betzl and Feist 2019)" review this paper},
}

@Report{Engborn2018,
  author      = {Engborn, Niklas and Moser, Christian},
  date        = {2018},
  institution = {Federal Reserve Bank of Minneapolis},
  title       = {Earnings Inequality and the Minimum Wage: Evidence from Brazil},
  type        = {resreport},
  doi         = {10.21034/iwp.7},
  url         = {https://www.minneapolisfed.org/research/institute-working-papers/earnings-inequality-and-the-minimum-wage-evidence-from-brazil},
}

@Article{Cengiz2019,
  author       = {Cengiz, Doruk and Dube, Arindrajit and Lindner, Attila and Zipperer, Ben},
  date         = {2019-05},
  journaltitle = {The Quarterly Journal of Economics},
  title        = {The Effect of Minimum Wages on Low-Wage Jobs},
  doi          = {10.1093/qje/qjz014},
  issn         = {0033-5533},
  number       = {3},
  pages        = {1405-1454},
  url          = {https://academic.oup.com/qje/article/134/3/1405/5484905},
  volume       = {134},
}

@Article{Dube2019,
  author       = {Arindrajit Dube},
  date         = {2019-10},
  journaltitle = {American Economic Journal: Applied Economics},
  title        = {Minimum Wages and the Distribution of Family Incomes},
  doi          = {10.1257/app.20170085},
  number       = {4},
  pages        = {268--304},
  url          = {https://www.aeaweb.org/articles?id=10.1257/app.20170085},
  volume       = {11},
  publisher    = {American Economic Association},
}

@Article{Harasztosi2019,
  author       = {Peter Harasztosi and Attila Lindner},
  date         = {2019-08},
  journaltitle = {American Economic Review},
  title        = {Who Pays for the Minimum Wage?},
  doi          = {10.1257/aer.20171445},
  issn         = {0002-8282},
  number       = {8},
  pages        = {2693--2727},
  volume       = {109},
  comment      = {Covers effects of minimum wage on employment, prices, and productivity.  Recommended in response to a submission to AEJ.},
  publisher    = {American Economic Association},
}

@Article{Dube2016,
  author       = {Arindrajit Dube and T. William Lester and Michael Reich},
  date         = {2016-07},
  journaltitle = {Journal of Labor Economics},
  title        = {Minimum Wage Shocks, Employment Flows, and Labor Market Frictions},
  doi          = {10.1086/685449},
  number       = {3},
  pages        = {663--704},
  volume       = {34},
  publisher    = {University of Chicago Press},
}

@Article{Gramlich1976,
  author       = {Eward M. Gramlich},
  date         = {1976},
  journaltitle = {Brooking Papers on Economic Activity},
  title        = {Impact of Minimum Wages on Other Wages, Employment, and Family Incomes},
  doi          = {10.2307/2534380},
  number       = {2},
  pages        = {409--462},
  url          = {https://www.brookings.edu/wp-content/uploads/1976/06/1976b_bpea_gramlich_flanagan_wachter.pdf},
  volume       = {7},
  publisher    = {{JSTOR}},
}

@Article{Burda2019,
  author       = {Michael C. Burda and Katie R. Genadek and Daniel S. Hamermesh},
  date         = {2019-10-02},
  journaltitle = {Economica},
  title        = {Unemployment and Effort at Work},
  doi          = {10.1111/ecca.12324},
  issue        = {347},
  pages        = {662--681},
  volume       = {87},
  publisher    = {Wiley},
  year         = {2019},
}

@Report{Mayer2004,
  author      = {Gerald Mayer},
  date        = {2004-08-31},
  institution = {Congressional Research Service},
  title       = {Union Membership Trends in the United States},
  type        = {resreport},
  location    = {Washington, DC},
  url         = {https://digitalcommons.ilr.cornell.edu/cgi/viewcontent.cgi?article=1176&context=key_workplace},
  urldate     = {2020-10-27},
}

@Report{Dube2014,
  author      = {Arindrajit Dube},
  institution = {Hamilton Project},
  title       = {Designing Thoughtful Minimum Wage Policy at the State and Local Levels},
  type        = {resreport},
  url         = {https://www.brookings.edu/wp-content/uploads/2016/06/state_local_minimum_wage_policy_dube.pdf},
  urldate     = {2020-09-27},
  month       = {6},
  year        = {2014},
}

@Article{Brummund2016,
  author       = {Peter Brummund and Michael R. Strain},
  date         = {2016-10-26},
  journaltitle = {Journal of Human Resources},
  title        = {Does Employment Respond Differently to Minimum Wage Increases in the Presence of Inflation Indexing?},
  doi          = {10.3368/jhr.55.2.1216.8404r2},
  number       = {3},
  pages        = {999--1024},
  volume       = {55},
  publisher    = {University of Wisconsin Press},
}

@Legislation{USCongress2019HR582,
  date     = {2019-07-22},
  location = {H.R 582, 116th Cong.},
  title    = {Raise the Wage Act},
  url      = {https://www.congress.gov/bill/116th-congress/house-bill/582},
}

@Legislation{USCongress2020HR2080,
  date     = {2020-04-04},
  location = {H.R 2080, 116th Cong.},
  title    = {PHASE Act},
  url      = {https://www.congress.gov/bill/116th-congress/house-bill/2080},
}

@Online{USSSACOLA2020,
  date         = {2020-01-01},
  title        = {Average Wages, Median Wages, and Wage Dispersion},
  url          = {https://www.ssa.gov/oact/cola/central.html},
  organization = {Social Security Administration},
  urldate      = {2020-01-01},
  year         = {2020},
}

@Online{EPIData2019,
  title        = {State of Working America Data Library},
  url          = {https://www.epi.org/data/},
  organization = {Economic Policy Institute},
  year         = {2019},
}

@Report{Schmitt2013,
  author      = {John Schmitt},
  institution = {Center for Economic and Policy Research},
  title       = {Why Does the Minimum Wage Have No Discernible Effect on Employment?},
  type        = {resreport},
  pagetotal   = {28},
  url         = {https://cepr.net/documents/publications/min-wage-2013-02.pdf},
  month       = {2},
  year        = {2013},
}

@Online{WorldBank,
  title        = {World Bank Data},
  url          = {https://data.worldbank.org},
  organization = {The World Bank},
  year         = {2020},
}

@Online{Krugman1997,
  author  = {Paul Krugman},
  date    = {1997-01-23},
  title   = {The Accidental Theorist},
  url     = {https://web.mit.edu/krugman/www/hotdog.html},
  urldate = {2020-10-30},
}
